\documentclass[10pt,twocolumn,letterpaper]{article}

\usepackage[pagenumbers]{cvpr} 

\usepackage{graphicx}
\usepackage{amsmath}
\usepackage{amssymb}
\usepackage{booktabs}

\usepackage[pagebackref,breaklinks,colorlinks]{hyperref}

\usepackage[capitalize]{cleveref}
\crefname{section}{Sec.}{Secs.}
\Crefname{section}{Section}{Sections}
\Crefname{table}{Table}{Tables}
\crefname{table}{Tab.}{Tabs.}

\def\cvprPaperID{*****} 
\def\confName{CVPR}
\def\confYear{2023}

\begin{document}

\title{Artificial Intelligence Security Competition (AISC)}

\author{Yinpeng Dong$^{1,2}$\thanks{Yinpeng Dong, Peng Chen, Senyou Deng and Lianji Li are organizers of this competition. The other authors are competitors of the top-ranking teams who contribute the solutions. The authors are ordered by the time of submission.}, Peng Chen$^1$, Senyou Deng$^1$, Lianji Li$^1$, Yi Sun$^3$, Hanyu Zhao$^3$, Jiaxing Li$^3$,\\ Yunteng Tan$^3$, Xinyu Liu$^3$, Yangyi Dong$^4$, Enhui Xu$^5$, Jincai Xu$^5$, Shu Xu$^5$, Xuelin Fu$^5$, \\ Changfeng Sun$^6$,  Haoliang Han$^6$, Xuchong Zhang$^6$, Shen Chen$^7$, Zhimin Sun$^7$, Junyi Cao$^7$, \\ Taiping Yao$^7$,  Shouhong Ding$^7$, Yu Wu$^8$, Jian Lin$^9$, Tianpeng Wu$^9$, Ye Wang$^8$, Yu Fu$^8$, Lin Feng$^{10}$, \\ Kangkang Gao$^{10}$,
Zeyu Liu$^{2}$, Yuanzhe Pang$^{2}$, Chengqi Duan$^{2}$, Huipeng Zhou$^{3}$, Yajie Wang$^{3}$, \\ Yuhang Zhao$^{3}$, Shangbo Wu$^{3}$, Haoran Lyu$^{3}$, Zhiyu Lin$^{11}$, Yifei Gao$^{11}$, Shuang Li$^{11}$, Haonan Wang$^{11}$, \\ Jitao Sang$^{11}$, Chen Ma$^{6}$, Junhao Zheng$^{6}$, Yijia Li$^{6}$, Chao Shen$^{6}$, Chenhao Lin$^{6}$, Zhichao Cui$^{12}$, \\ Guoshuai Liu$^{12}$, Huafeng Shi$^{12}$, Kun Hu$^{12}$, Mengxin Zhang$^{12}$
\vspace{0.3cm}\\
$^{1}$ RealAI \; $^{2}$ Tsinghua University \; $^{3}$ Beijing Institute of Technology \; $^{4}$ Shanghai Jiao Tong University \\ $^{5}$ China Nanhu Academy of Electronics and Information Technology \;
$^{6}$ Xi’an Jiaotong University  \\ $^{7}$ Tencent YouTu Lab \; $^{8}$ China Construction Bank Fintech \; $^{9}$ RippleInfo\\
$^{10}$ Zhejiang Dahuatech Co., Ltd. \; $^{11}$ Beijing Jiaotong University \; $^{12}$ Xidian University
}

\maketitle


\begin{abstract}
    The security of artificial intelligence (AI) is an important research area towards safe, reliable, and trustworthy AI systems. To accelerate the research on AI security, the Artificial Intelligence Security Competition (AISC) was organized by the Zhongguancun Laboratory, China Industrial Control Systems Cyber Emergency Response Team, Institute for Artificial Intelligence, Tsinghua University, and RealAI as part of the Zhongguancun International Frontier Technology Innovation Competition (\url{https://www.zgc-aisc.com/en}). The competition consists of three tracks, including Deepfake Security Competition, Autonomous Driving Security Competition, and Face Recognition Security Competition. This report will introduce the competition rules of these three tracks and the solutions of top-ranking teams in each track.
\end{abstract}

\section{Introduction}

With the rapid development of deep learning algorithms \cite{DeepReview_Lecun_2015}, they have been widely applied to numerous applications, such as face recognition, autonomous driving, \etc. However, researchers have found that these models have security issues in many aspects. As a typical example, deep neural networks are vulnerable to the maliciously generated adversarial examples~\cite{szegedy2013intriguing,goodfellow2014explaining,Dong_2018_CVPR}, which can mislead the models by adding small perturbations to normal ones. Adversarial examples raise great concerns about the safety and reliability of deep learning in real-world applications, such as face recognition, autonomous driving. Besides, deep learning techniques can be utilized for malicious purposes, such as generating fake human faces, known as Deepfake. 

To accelerate the research on AI security, we organized the Artificial Intelligence Security Competition (AISC). This competition consists of three tracks, including Deepfake Security Competition (Sec.~\ref{sec:2}), Autonomous Driving Security Competition (Sec.~\ref{sec:3}), and Face Recognition Security Competition (Sec.~\ref{sec:4}). Below we introduce the competition details and top-ranking submissions of each track.

\section{Deepfake Security Competition}
\label{sec:2}

This track focuses on the governance of AI-generated content, also known as Deepfake. With the continuous emergence of easy-to-use face editing and vocal simulation tools, the malicious abuse of Deepfake technology has caused public concern. For example, criminals use this technology to synthesize audio and video content of specific person for telecommunication fraud, fabricating fake news, etc., causing serious negative impacts on individuals and society. Therefore, lots of works have been proposed to address the real/fake detection task~\cite{zhao2021multi,sun2022dual,haliassos2022leveraging} and Deepfake source identification~\cite{yu2019attributing,jia2022model,girish2021towards} task. 

It is one of the promising countermeasures to identify the manipulation methods hidden behind forged contents by analyzing the similarity between them. Thus, this competition defines a novel task to solve Deepfake source identification in open-world scenarios. Given a query image, identify the Deepfake method behind it based on its similarities to the images in the gallery set.

\subsection{Competition Rules}

This track is divided into three stages: preliminary competition, preliminary evaluation and final competition.

\textbf{Preliminary competition:} This phase examines the generalization of the semi-supervised Deepfake method identification algorithms, including their generalization to different source images and post-processing methods. The provided training dataset consists of two parts: the labelled dataset $\mathcal{D}_l^{tr}$and the unlabelled dataset $\mathcal{D}_u^{tr}$. To represent data with informed Deepfake synthesis methods, each sample in the labelled dataset $\mathcal{D}_l^{tr}$provides information of its source images and deepfake synthesis method. The unlabelled dataset $\mathcal{D}_u^{tr}$ does not provide any relevant information, which represents forged data that are collected in the wild (e.g., from social media, websites) with uninformed deepake synthesis methods and source images. The Deepfake synthesis methods used in the labelled dataset are clearly defined, and we denote the set of these methods as $\mathcal{Y}_l$. The unlabelled dataset contains both the defined Deepfake methods $\mathcal{Y}_l$ and some undefined Deepfake methods $\mathcal{Y}_u$, where$\mathcal{Y}_l\cap\mathcal{Y}_u = \emptyset$. The undefined methods represent methods that are unknown to or not simulated by the developers who train the identification algorithms. Note that these undefined Deepfake methods may exhibit data imbalance phenomenon on their synthetic samples. Denote the complete set of Deepfake synthesis methods in the training set as $\mathcal{Y}_{tr}\triangleq\mathcal{Y}_{l}\cup\mathcal{Y}_u$. Participants can also use additional private datasets and public datasets under restrictions to train their models. The forged samples in the private dataset can only be manipulated by methods from$\mathcal{Y}_l$or their simple variants (e.g., modifying pre-processing and post-processing steps, re-training the models, combining the methods in sequence, etc.). 

The testing process introduces a probe set $\mathcal{D}^{p}$ and a gallery set $\mathcal{D}^{g}$. All Deepfake methods used in $\mathcal{D}^{p}$ and $\mathcal{D}^{g}$are from $\mathcal{Y}_{tr}$. Participants need to submit the inference programs of their models and the corresponding docker runtime environments for online evaluation. Participants' algorithm should be able to process the hidden probe set $\mathcal{D}^{p}$and the gallery set $\mathcal{D}^{g}$. For each sample in the probe set $\mathcal{D}^{p}$, the program returns the five samples with the highest similarities in the reference set $\mathcal{D}^{g}$and their similarities. The testing process examines the generalization of the submitted retrieval algorithms. Therefore, the appearance and post-processing methods of the test data will be somewhat different from the training data.

\begin{table*}[!t]
\centering
\caption{The quantitative results of the final top-5 teams of Deepfake security competition.}
\begin{tabular}{c|cccc}
\toprule
Team Name                           & Precision@5 & AUC  & Subjective score  & Total  \\
\midrule
AreYouFake                          & 0.9820        & 0.9944 & 1             & 0.9875   \\
hello world                         & 0.9742        & 0.9784 & 0.9             & 0.9680   \\
Forgery identification right?       & 0.8927        & 0.8822 & 0.85             & 0.8850   \\
CanCanNeed                          & 0.8673        & 0.9165 & 0.85             & 0.8803   \\
TianQuan $\&$ DaHua & 0.8708        & 0.8906 & 0.9             & 0.8796   \\
\bottomrule
\end{tabular}
\label{tab:Deepfake-top-5-results}
\end{table*}

\begin{figure*}[!t]
    \centering
    \includegraphics[width=\linewidth]{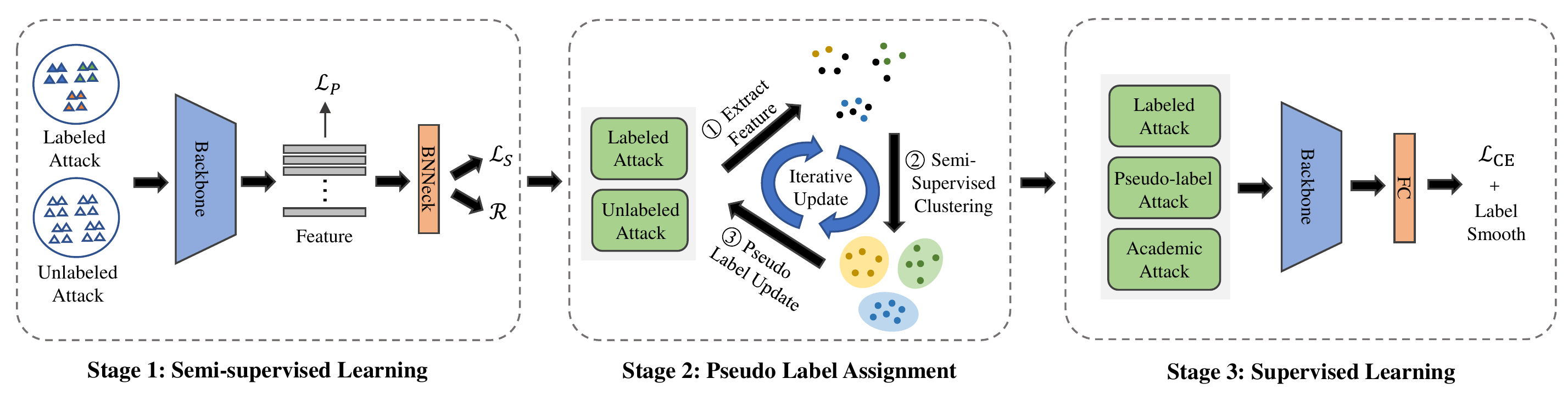}
    \caption{The proposed Multi-stage Semi-supervised Deepfake Attribution (MSDA) framework.}
    \label{fig:network}
\end{figure*}

\textbf{Preliminary evaluation:} This phase examines the generalization performance of the Deepfake method identification algorithms on new Deepfake methods. Participants' last submitted programs at the end of the preliminary competition will be evaluated using a different probe set and gallery set. In this phase, the Deepfake methods used in the probe set and the gallery set are not only from $\mathcal{Y}_{tr}$, but also from a hidden Deepfake method set $\mathcal{Y}_h$, where $\mathcal{Y}_{tr}\cap\mathcal{Y}_h = \emptyset$. The scoring for the preliminary evaluation is the same as that for the preliminary competition. After the evaluation, the organizing committee will contact the top-ranked teams and select the 10 teams with the highest scores to enter the final competition. 

\textbf{Final competition:} In this phase, we also examine Deepfake method identification algorithms' ability to discover novel Deepfake methods online. The requirements for the training set in the final competition are the same as those in the preliminary competition, and the reference set in the testing process is the same as in the preliminary evaluation. Given the query set from the preliminary evaluation, the query set used in the final competition will add the data generated by the new Deepfake methods $y\in\mathcal{Y}_{new}$, where $\mathcal{Y}_{new}\cap(\mathcal{Y}_{tr} \cup\mathcal{Y}_{h}) = \emptyset$. For each data in the query set, the algorithm also needs to predict the probability$p\in[0, 1]$ that the data is generated by the new Deepfake method. At the same time, participants are required to submit a description of their algorithm solutions so that the experts of the organizing committee can evaluate and score the participants' solutions subjectively. 

\textbf{Evaluation Metric:}
In the preliminary competition and preliminary evaluation, submissions are evaluated according to top-5 precision:
\begin{equation}
    precision@5 =\frac{1}{5N} \sum_{i=1}^{N} \sum_{j=1}^{5} I(\hat y_{i,j} = y_i),
\end{equation}
where:
\begin{itemize}
    \item $N$ is the number of images in  probe set $\mathcal{D}^{p}$;
    \item $y_i$ is label of the i-th image in query set $\mathcal{D}^{p}$;
    \item $\hat y_{i,j}$ is label of the ranked j-th image recalled from the gallery set $\mathcal{D}^{g}$ for the i-th image in the probe set $\mathcal{D}^{p}$;
    \item $I$ denotes the indicator function, return 1 if $y_i = \hat y_{i,j}$, otherwise, 0
\end{itemize}
In the final competition, the score is calculated based on the following sections:
\begin{itemize}
    \item For the data generated by Deepfake method in $\mathcal{Y}_{tr} \cup \mathcal{Y}_{h}$, calculate its precision@5 (same as the preliminary competition and preliminary evaluation);
    \item For all data in the probe set $\mathcal{D}^{p}$, calculate the AUC (Area under the ROC Curve) according to the predicted probability of belonging to the new forgery method;
    \item The experts of the organizing committee will subjectively assess the submitted solutions and score them in terms of clarity of presentation, reproducibility, and research value.
\end{itemize}
Thus, the final score is computed by:
\begin{equation}
   S_{final} = 0.6 \times precision@5  + 0.3 \times AUC + 0.1 \times S_{sbj}.
\end{equation}

\subsection{Competition Results}

In this competition track, there are 144 teams participate in the preliminary competition, and submit results 912 times totally. The quantitative results of final top-5 teams are shown in Table \ref{tab:Deepfake-top-5-results}.

\subsection{Top Scoring Submissions}

\subsubsection{1st place:  AreYouFake-Tencent\_YouTu}

\begin{center}
    \textbf{Team Member}
\end{center} 

Shen Chen, Zhimin Sun, Junyi Cao, Taiping Yao, Shouhong Ding

\begin{center}
    \textbf{Method}
\end{center}

For the Deepfake attribution task in the preliminary, they designed a \textbf{Multi-stage Semi-supervised Deepfake Attribution (MSDA)} framework that contains three stages, namely semi-supervised learning, pseudo label assignment and supervised learning, as shown in Figure~\ref{fig:network}. The details of each stage are as follows:

\textbf{Stage 1: Semi-supervised Learning.}
For inputs that contain both labeled and unlabeled attacks, they use the pairwise similarity of features to facilitate automatic clustering of the same type of attacks, while introducing a category entropy maximization rule~\cite{cao2021open} to enhance the difference between different attacks. Besides, BNNeck~\cite{luo2019strong} is introduced to mitigate the effect of different loss constraints on feature learning. With the above techniques, they can make full use of unlabeled data while balancing the learning process between known and unknown attacks.

\textbf{Stage 2: Pseudo Label Assignment.}
They use the model trained in stage 1 to extract features of both labeled and unlabeled data, and then perform clustering by Semi-supervised K-Means algorithm~\cite{vaze2022generalized} and adaptively assign corresponding pseudo-labels to known attacks and unknown attacks on unlabeled data. With the assigned pseudo-labels, they can directly perform supervised learning without additional constraints. The entire process of stage 2 is repeated several times to obtain the pseudo-labels with higher confidence.

\textbf{Stage 3: Supervised Learning.}
Based on the pseudo-labels obtained in stage 2, they introduce academic datasets and perform directly supervised learning to improve the retrieval performance on more types of attacks. To enhance generalizability, they also employ strategies such as label smoothing, data augmentation and model ensemble.

\begin{figure*}[t]
    \centering
    \includegraphics[width=\linewidth]{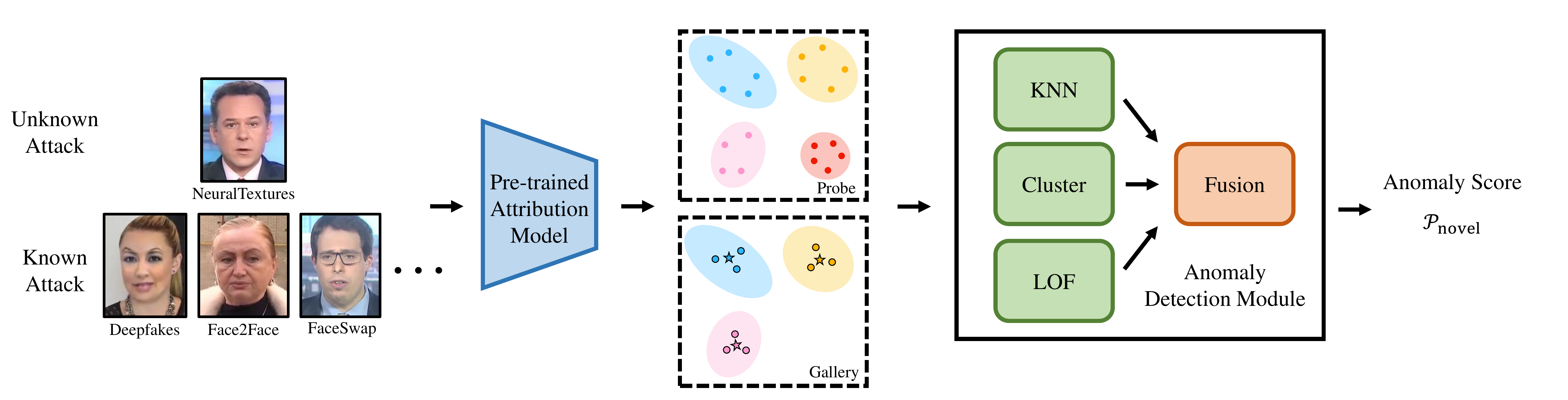}
    \caption{The proposed Multi-module Anomaly Detection (MAD) framework.}
    \label{fig:novel_detection_framework}
\end{figure*}

\begin{table*}[h]
    \centering
    \caption{Top 5 accuracy of different experimental strategies on the online preliminary evaluation set.}
    \begin{tabular}{c|c|c|l|l|c}
    \hline
    \textbf{Experiment} & \textbf{Backbone} & \textbf{Size} & \textbf{Training Set} & \textbf{Training Strategy} & \textbf{Top-5 Accuracy} \\ \hline
    (1) & EN-B0 & 224 & DFSC & Five Classification & 0.914090 \\ \hline
    (2) & EN-B0 & 224 & DFSC & Semi-Supervised Learning & 0.921003 \\ \hline
    (3) & EN-B4 & 320 & DFSC & Semi-Supervised Learning & 0.933916 \\ \hline
    (4) & EN-B4 & 320 & DFSC & (3) + BNNeck & 0.955640 \\ \hline
    (5) & EN-B4 & 320 & DFSC & (4) + Pseudo-label assignment & 0.971212 \\ \hline
    (6) & EN-B4 & 320 & DFSC & (5) + Crop Face & 0.987391 \\ \hline
    (7) & EN-B4 & 320 & DFSC + Academic Sets & (6) + Supervised Learning & 0.988606 \\ \hline
    (8) & EN-B4 & 320 & DFSC + Academic Sets & (7) + Model Ensemble & 0.994379 \\ \hline
    \end{tabular}
    \label{tab:results}
\end{table*}

\begin{table*}[h]
     \centering
     \caption{Quantitative results of different anomaly detection modules on the local evaluation set.}
     \begin{tabular}{c|c|c|c|c|c}
         \hline
         \textbf{Experiment} & \textbf{Model} & \textbf{Anomaly Module} & \textbf{Top-5 Accuracy} & \textbf{AUC} & \textbf{Time Consuming}\\
         \hline
         (a) & Experiment (7) & KNN & 0.967501 & 0.967310 & 220s \\
         \hline
         (b) & Experiment (7) & LOF & 0.967501 & 0.968589 & 328s \\
         \hline
         (c) & Experiment (7) & Cluster & 0.967501 & 0.971290 & 242s \\
         \hline
         (d) & Experiment (8) & KNN & 0.994851 & 0.984262 & 660s \\
         \hline
         (e) & Experiment (8) & LOF & 0.994851 & 0.986549 & 971s \\
         \hline
         (f) & Experiment (8) & Cluster &0.994851 & 0.990264 & 711s \\
         \hline
     \end{tabular}
     \label{tab:novelty_results}
\end{table*}

For the anomaly detection task in the final, they proposed a \textbf{Multi-module Anomaly Detection (MAD)} framework, which includes three anomaly detection modules based on K-nearest neighbors, cluster and local outlier factor, respectively. The overall anomaly detection algorithm flow is shown in Figure~\ref{fig:novel_detection_framework}, the details are as follows:

\textbf{K-nearest neighbors (KNN)-based anomaly detection.}
The feature distance between each probe sample and all gallery samples is first calculated, and then the average of the K nearest neighbor distances is taken as the anomaly score. For known attacks, there are at least K samples with the same attack type in the gallery set; for novel attacks, there are no samples with the same attack type in the gallery set. Therefore, anomaly detection can be achieved by simply calculating the average of the top-k distances.

\textbf{Cluster-based anomaly detection.}
Compared with the K-nearest neighbor scheme, this module is more robust to noise because it considers the clustering centers in the probe and gallery. We first cluster the sample features in probe and gallery separately by the KMeans algorithm to obtain $p$ probe clusters and $g$ gallery clusters. For each probe sample, its novel score consists of two parts. The first part is the closest distance between the probe feature and the center of $g$ gallery clusters. The second part is the closest distance between the cluster center to which the probe sample belongs and the center of $g$ gallery clusters. The weights of these two components in the overall anomaly score are $w_1$, $w_2$, respectively.

\textbf{Local Outlier Factor (LOF)-based anomaly detection.}
LOF measures the local deviation of the density of a given sample concerning its neighbors. It is local in that the anomaly score depends on how isolated the object is concerning the surrounding neighborhood. More precisely, locality is given by k-nearest neighbors, whose distance is used to estimate the local density. By comparing the local density of a sample to the local densities of its neighbors, one can identify samples that have a substantially lower density than their neighbors, namely novel attack samples.

For a given probe sample, they calculate the anomaly scores through the above three modules, and then fuse the three scores to obtain the final anomaly score.

\begin{center}
    \textbf{Submission Details and Results}
\end{center}

\textbf{Datasets.}
The datasets they use in this competition mainly include the officially provided Deepfakes Security Challenge (DFSC) dataset, as well as five academic datasets, \textit{i.e.} FaceForensics++~\cite{rossler2018faceforensics}, CelebDF~\cite{li2019celeb}, DeeperForensics-1.0~\cite{jiang2020deeperforensics}, ForgeryNet~\cite{he2021forgerynet} and FakeAVCeleb~\cite{khalid2021fakeavceleb}. Considering the large differences in the distribution of different attack types, they implemented different sampling strategies for different datasets to maintain the class balance.

\textbf{Implementation Details.}
The team implements the proposed method via PyTorch. All the models are trained on 4 NVIDIA Tesla V100 GPUs. They use the EfficientNet-b4~\cite{tan2019efficientnet} pre-trained on ImageNet~\cite{deng2009imagenet} as the backbone. The input image size is scaled to $320 \times 320$. Adam~\cite{kingma2014adam} is used as the optimizer with a learning rate of 0.0002. A total of 50 epochs are trained and the learning rate decreases to the original 0.2 every 10 epochs. They use dlib\footnote{https://github.com/davisking/dlib} as the face detector and expand the region by 1.2 times to include more facial information. Data augmentations such as random horizontal flip, random brightness change, and random cropping were used. During inference time, random horizontal flip was used as Test Time Augmentation (TTA) to further improve model performance.

\textbf{Preliminary Results.}
For the Deepfake attribution task, the team lists the Top-5 accuracies of different models and training strategies on the online preliminary evaluation set in Table~\ref{tab:results}. Comparing experiments (1) and (2), the accuracy improves by 0.70\% which mainly benefits from semi-supervised learning, and the performance further improves to 0.933916 after using a large model and large image size. In experiment (4) they introduce BNNeck~\cite{luo2019strong}, in which the Top-5 Accuracy has been significantly improved. Then experiment (5) illustrates the effectiveness of the pseudo-label assignment strategy in stage 2. Finally, they introduced the academic dataset and used model ensemble as the final strategy to achieve the best result of 0.994379.

\textbf{Final Results.}
For the anomaly detection task, they use DFSC and academic datasets to construct a local evaluation set to verify the performance of different schemes. Then, the pre-trained models in the Deepfake attribution task were used as the feature extractors. Specifically, the models trained on experiment (7) and experiment (8) were used in the single-model setting and the multi-model setting, respectively.
Table~\ref{tab:novelty_results} shows the quantitative results of different anomaly detection modules on the local evaluation set, including top-5 accuracy, AUC and time cost.
Comparing experiments (a), (b) and (c), they find that the Cluster-based module achieves the best AUC result, followed by the LOF-based and KNN-based, but LOF usually cost longer time than the other modules. Then they repeat the experiments with the multi-model setting, resulting in better results. They finally combine those three scores and obtain the final anomaly score.

\subsubsection{2nd place: hello world}

\begin{center}
    \textbf{Team Member}
\end{center}

Zhichao Cui, Guoshuai Liu, Huafeng Shi, Kun Hu, Mengxin Zhang

\begin{center}
    \textbf{Method}
\end{center}

Since there are both known types of data and unknown types of data in the training data set, we adopt a supervised classification framework and a scheme of pseudo-labeling unknown types of data to solve the preliminary task. In the abnormal sample detection task, we obtain the abnormal sample probability by K-means of clustering and entropy probability.

Baseline: Collect the given four types of training data, as well as 5 public academic set data, and use Resnest50 as the backbone to train a classification model. In order to prevent the occurrence of category imbalance, we sample about 2k samples for each category of data, a total of 22 categories, training images of more than 5.5W, and then obtain a baseline model.

Unknown type data prediction: Mark as a known category sample. The first step is to infer the scores of all unknown category data, and sort them, take the top 90\% of the samples, and filter out the categories with less than 10 samples to obtain 13 known category samples; The second step, filter out the scores less than 0.7. Finally, 2146 pseudo-label data were obtained. Add the labeled data to training to get model-1. Mark as a new class sample. The first step, model-1 is used to extract the remaining unknown category sample features; The second step, K-means clustering is used for the above features, and the k value is 3 according to the T-SNE feature distribution visualization result (Figure~\ref{fig:top2_kmeans}) to get model-2:

\begin{figure}
\centering
\includegraphics[width=0.9\linewidth]{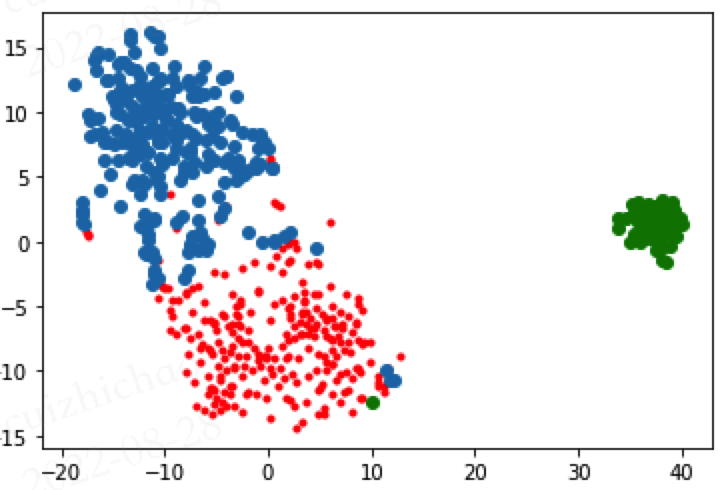}
\caption{ T-SNE visualization results}    
\label{fig:top2_kmeans}
\end{figure}

Ensemble: The knowledge learned by different models has certain differences, especially under different backbones. Integrating different models will give better performance than a single model. Here, ResNest 50, ResNest 101, and efficient b4 are selected for integration, and the cosine similarity is calculated after concatenating multiple model features to obtain the final score.

Similar to how model-2 predicts labels of unknown types of data, we relabel pseudo-labels with features and train with arcface loss, obtain model-3, and use their concatenated features for anomaly detection tasks.

Anomaly detection: Perform kmeans clustering on the features of the gallery data, and according to the data prior (at least 5 of the same type of data exist in the gallery), we set the number of clusters to be $num_g$ / 5, and then calculate the cluster center features of each center. Finally, calculate the cosine similarity between the probe data and all centerfeas, take the highest similarity value to calculate the entropy, and obtain the abnormal sample probability value. Calculated as follows:
\begin{eqnarray}
    SIM = F_P^T F_C \frac{1}{\Vert F_P\Vert\Vert F_C\Vert}\\
    P = - SIM \log SIM
\end{eqnarray}
where $F_P$ is the feature of the probe, $F_C$ is the central feature, and SIM is the cosine similarity.

\begin{center}
    \textbf{Submission Details and Results}
\end{center}

The input image size is scaled to $320 \times 320$. The SGD is used as the optimizer with a initial learning rate of 0.02. And the learning rate decreases to the original 0.1 every 2,000 steps. The preliminary score is 0.97, the final score is 0.96.

\subsubsection{3nd place: Forgery identification right?}

\begin{center}
    \textbf{Team Member}
\end{center}

Sun Yi, Zhao HanYu, Li JiaXing, Tan YunTeng, Liu XinYu

\begin{center}
    \textbf{Method}
\end{center}

\begin{figure}[t]
 \centering
 \includegraphics[scale=0.30]{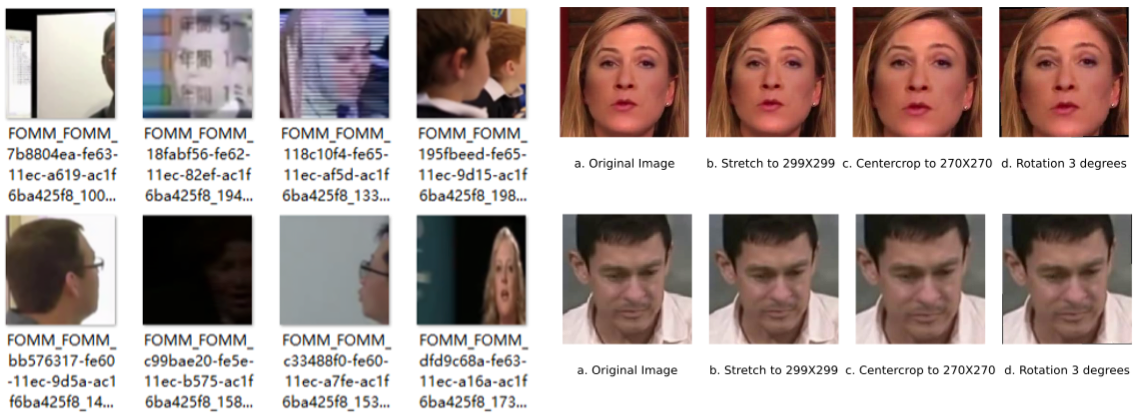}
 \caption{Data preprocessing and augmentation. Left: image culled by automatic data cleaning. Right: the data is augmented to reduce postprocessing variance of Deepfake samples and prevent overfitting.}
\label{fig1}
\end{figure}

\textbf{Data Cleaning and Augmentation.}
The first step is to drop out data that does not contain valid fake faces. After uniformly scaling each image to 299x299 size, crop to 270x270-299x299 using a random centre, use random skew from -3 degrees to +3 degrees and stretch to 299x299, as shown in Figure~\ref{fig1}. Data were normalized to a three-channel mean and variance of 0.5. Processing Deepfake image data in this way has two advantages. One is that it can give samples more consistent post-processing. In general, the Deepfake recognition system is sensitive to high-frequency information. There will be certain post-processing operations in all aspects of Deepfake production and dissemination, so the post-processing operations of different datasets are approximate. These enhancements can make the model more focused on Deepfake. The difference in methods is not the difference in post-processing of deep fake samples; the second is because the amount of data given by the competition is too small; by increasing the number and diversity of examples, the model can be prevented from overfitting.

\textbf{The Unlabeled Data.}
The trained network can recognize Deepfake images, indicating that the network can learn the characteristics of these forgery methods. We first use these unlabeled data for true-false binary classification pre-training and use the Xception network as the feature extractor, as shown in Figure~\ref{fig3}.
There is a specific correlation between various forgery methods and similar generation processes. Drawing on the pre-training idea of Contrastive Learning, use the deep fake label data in the officially provided public data set to train a multi-class model, so that the model can learn the characteristics of various Deepfake methods in depth. Then use the pre-training model to extract the features of the None set, and obtain the pseudo labels of the None set by clustering, as shown in Figure~\ref{fig4}.

\begin{figure}[t]
 \centering
 \includegraphics[scale=0.25]{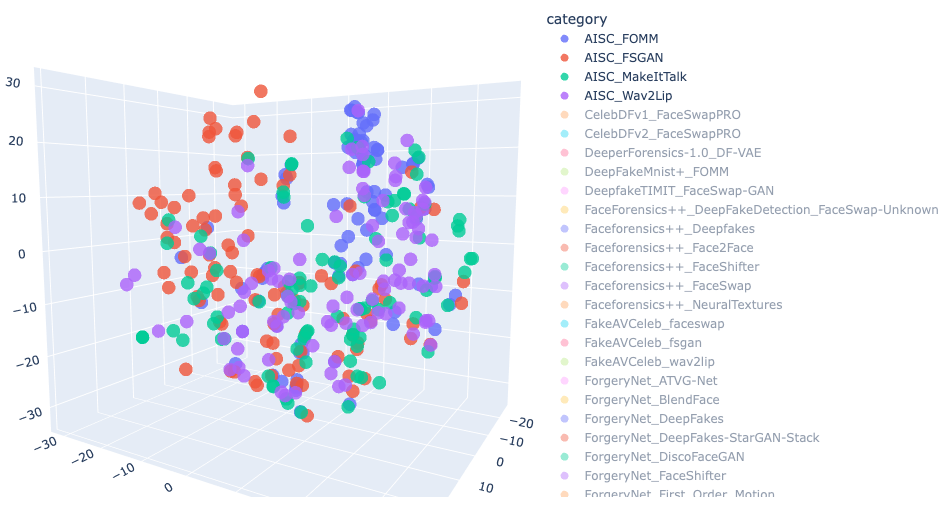}
 \caption{Scatter plots of FOMM, FSGAN, MakeItTalk, and Wav2Lip after training with unlabeled data. The results of the online evaluation also show that the technique improves the score from 0.14 to 0.36, proving the method is effective.}
\label{fig2}
\end{figure}

\begin{figure}[t]
 \centering
 \includegraphics[scale=0.25]{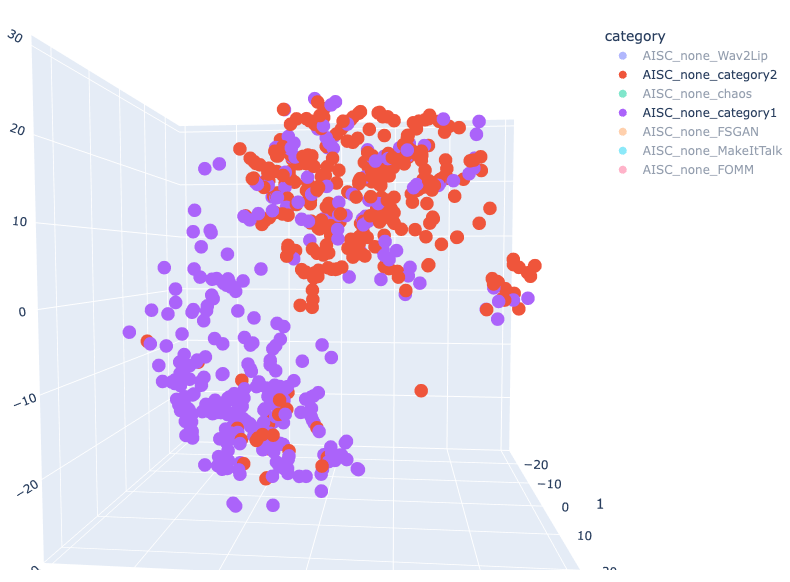}
 \caption{After removing the data labelled as FOMM, FSGAN, MakeItTalk, and Wav2Lip in the None Sets, cluster the remaining data in the none set into category1 and category2.}
\label{fig3}
\end{figure}

\textbf{Merge Similar Forgery Data.}
We used the cosine distance for various forgery methods to calculate the similarity matrix for each forgery method, using the pre-trained Xception model mentioned earlier as a feature extractor, as shown in Figure~\ref{fig4}. It is obvious that there are significant differences in the similarity between different forgery methods, and merging the methods with closer distances will be beneficial to improve the model performance.
Then perform multi-class training according to the forgery method indicated by each data set. The model is Xception neural network for 12 rounds based on the pre-training model. The initial learning rate is 0.001, and the learning rate decay is 50\% every 5 rounds. The model has an excellent distinguishing effect, as shown in Figure~\ref{fig5}.

\begin{figure}[t]
 \centering
 \includegraphics[scale=0.4]{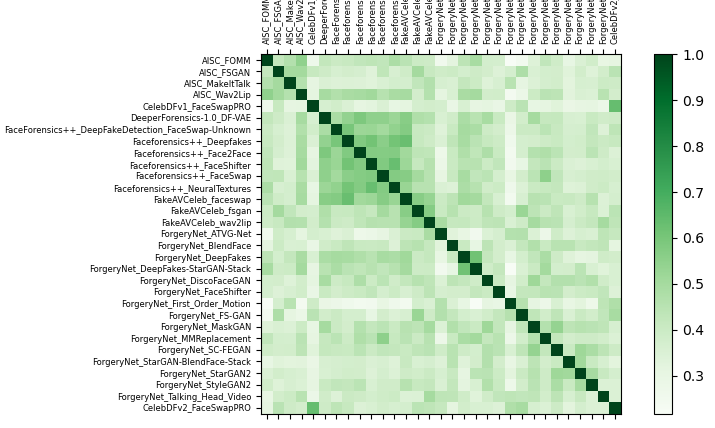}
 \caption{Similarity matrix of various forgery methods.}
\label{fig4}
\end{figure}

\begin{figure}[t]
 \centering
 \includegraphics[scale=0.25]{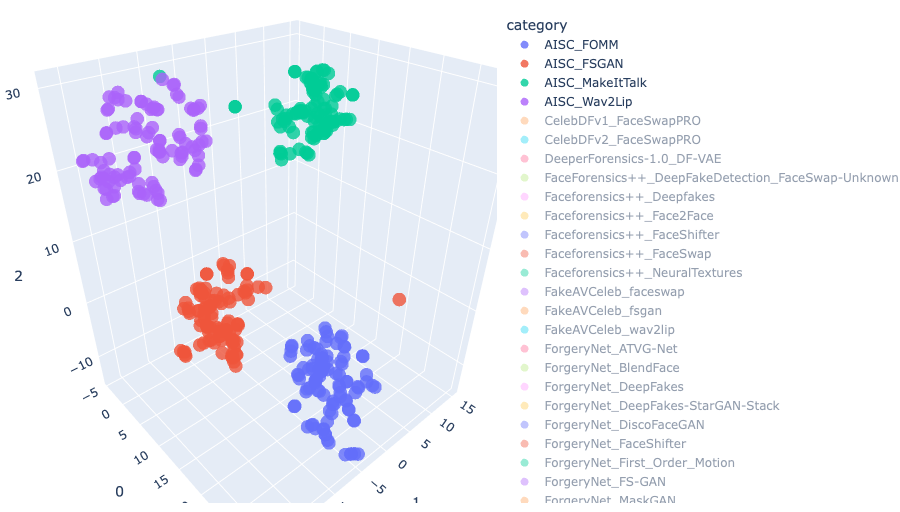}
 \caption{3D scatter plot of FOMM, FSGAN, MakeItTalk, and Wav2Lip after training with labelled data.}
\label{fig5}
\end{figure}

\subsubsection{4th place: CanCanNeed}

\begin{center}
    \textbf{Team Member}
\end{center}

Yangyi Dong

\begin{center}
    \textbf{Method}
\end{center}

\textbf{Data Standardization.} It is observed that some images in the training data look close to black. By observing the pixel values of these images, we find that some of the maximum and minimum pixel values are only 50, rather than 255. Therefore, the input images are normalized and then transmitted to the training network.

\textbf{Split of training set and test set.} During training, the training set and test set are divided by a ratio of 9:1. After super parameters are selected, all training data are used for training model parameters.

\textbf{External academic data.} In the external data, we tried FaceForensics \cite{rossler2018faceforensics}, DeeperForensics \cite{jiang2020deeperforensics}, ForgeryNet \cite{he2021forgerynet} and other data sets. Among them, FaceForensics will improve the indicators on the test set when training the eight category model, while other data sets have no effect.

\textbf{Unlabelled data processing.} Since there is no good method to label the unlabeled data, the unlabeled data are labelled as the fifth category to train the classification model.

\textbf{Network framework and loss function.} We use efficientnet-b4 as the backbone and cross entropy loss.

\begin{center}
    \textbf{Submission Details and Results}
\end{center}

At the beginning, we do not use unlabeled data. We only use labeled data to train a four classification network. The result of this method on test data is 0.77, which can be regarded as the baseline. After adding data standardization, the result increases from 0.77 to 0.78. After adding FaceForensics as training data, the result increases from 0.77 to 0.80. After labeling unlabeled data as a new category, the result increases from 0.80 to 0.88.

\subsubsection{5th place: TianQuan $\&$ DaHua}

\begin{center}
    \textbf{Team Member}
\end{center}

Lin Feng, Yu Wu, Kangkang Gao, Ye Wang and Yu Fu

\begin{center}
    \textbf{Method}
\end{center}

We improve the generalization of network for extracting better features from three aspects:
(1) enrich the composition of training set: we built a self-organized data set within the allowable range of the competition (some public data set and reproduce the allowable generation algorithm to generate data). We used data augmentation, image compression with a small probability (1\%) during data loading, and image post-processing with superimposed noise; 
(2) improve feature extraction capability: we selected a network with strong feature extraction capability, and designed the loss function to enhance the constraint of feature level; 
(3) developed three-stage training strategy: the unlabeled data was made full use of in the study.

\textbf{Backbone and loss function:}
Xception~\cite{chollet2017xception} was selected as our feature extraction backbone with 2048-dimension data, which is an effective classification network in the field of deep-fake. 
The loss function used cross entropy and center loss~\cite{wen2016discriminative}, with a ratio of 1:1. The cross entropy is used to constrain the predicted label of the network consistent with the actual label, and the center loss is used to constrain the within class distance in the feature dimension.

\textbf{Usage of unlabeled data:} 
Firstly, the classifier is trained with four types of officially labeled data, where the unlabeled data is no-used in preliminary. We further used unsupervised spectral clustering algorithm to classify the remaining unlabeled data. Finally, we combined other labeled data for re-training the  the classifier. In this stage, the label of unlabeled data will constantly update until the label is stable.

\begin{center}
    \textbf{Submission Details and Results}
\end{center}

There are total 18 types of deep-fake data used in our experiments: 4 types of officially labeled data, 7 types of unlabeled clustering, and 7 types of public data sets. In the data loading stage, the image is uniformly bi-cubic interpolated to a uniform size, and data enhancement operations such as inversion are performed for some categories with less data. In the final training stage, the initial learning rate is set to 0.001, and it is reduced 10\% every 20 epochs. The training step will be stopped when the learning rate drops to 0.000001. The score is 0.87 in the preliminary session. The final score is 0.88, which is the fifth in the final.

\section{Autonomous Driving Security Competition}
\label{sec:3}
This track focuses on the security of detection model in autonomous driving tasks. At present, the autonomous driving is booming  in industry field, however its structural imperfection also result in some security threats in auto-driving system. In 2016, one Tesla car drived by auto-driving system caused the first fatal traffic accident because this car failure to detect a white truck under strong sunlight. In a normal and safe environment, the probability of such extreme situations is extremely low, but the scary thing is that attackers can use algorithmic flaws to fool the perception system of auto-driving system to replicate such situations. For example, attackers will deceive the object detection model of auto-driving sytem by modifying the texture features of vehicles to make traffic accidents.

In order to better evaluate the security of object detection models in auto-driving system, this competition simulated the adversarial attack in autonomous driving environments for contestants to make experiments. We aim to attract more and more people to focus on the security of auto-driving system and make some meaningful validation experiments with contestants for further research on adversarial attack in auto-driving field in real world.

\subsection{Competition Rules}

This track is dividied to three stages: preliminary round, judgment round and final round. 
	
In the preliminary round, we will provide 10 simulated videos, where 5 videos have been sticked with a example patch and others are clean in the corresponding scene. these videos can been used as train or validation dataset. When contestants upload their patches to our scoring system, we will evaluated their results in another 5 different scenes simulated by Carla\cite{Dosovitskiy17} with a white-box (YOLOv3) and a black-box detection model. In each scene, the victim car will run normally and the attacked truck with participant's path will stop in front of it. The scoring system will count the number of frame which don't have target classes by above detection model to compute the score. Finally, the total score of one team is the average in 5 scenes, and the formula is summarized as:
\begin{equation}
	\begin{aligned}
		score=\frac{1}{2}\sum_{i=1}^{2}\frac{1}{5} \sum_{j=1}^{5} \frac{1}{240} \sum_{f_{j}=1 }^{240} I(M_{i}(x_{f_{j}})=0) \\
		 +  0.2*(1-\frac{\left \| m/255 \right \|_{1}}{2790*1260} )
	\end{aligned}
\end{equation}
where $I$ is a indicator function, $M_{i}$ denotes the $i$-th object detection model, $x_{f_{j}}$ denotes the No. $f_{j}$ frame extracted from the $j$-th scene, and $M_{i}(x_{f_{j}})$ denotes the number of "car", "truck" and "bus" detected by white or black detection model. If $M_{i}(x_{f_{j}})=0$, then the adversarial patch is work in this frame. $240$ is the number of frame of each ecene video, and $2790*1260$ is the size of a patch. The second item will compute the utilize acreage of one patch, and smaller acreage will obtain higher score. Additionally, in evaluation stage, the maximum limitation of the number of connected regions in a patch is 5, if the counted number is exceeded, this result will be void.

In the judgment round, we will check contestants' technical solutions and source code. If they are exactly the same to one of the open source resources, we will disqualify the team. After this, we adopted another black-box detection model to evaluate players' patches again and then chosen the top-10 teams to enter the final round.

\begin{figure}[h]
	\centering
	\resizebox*{0.5\textwidth}{!}{
		\begin{tabular}{cc}
			\includegraphics[scale=0.4]{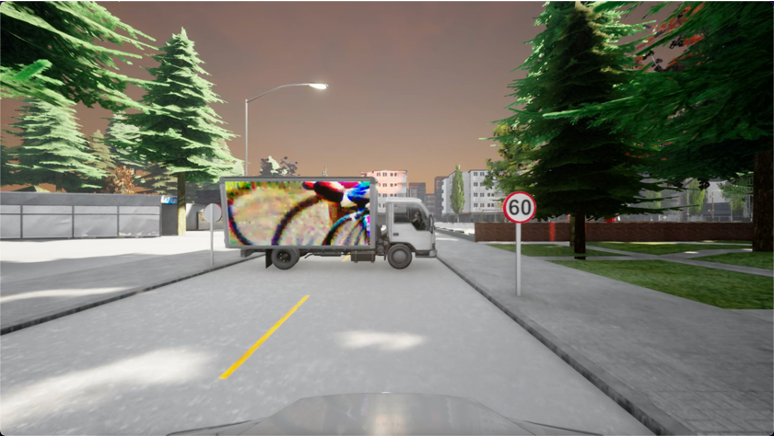} &
			\includegraphics[scale=0.4]{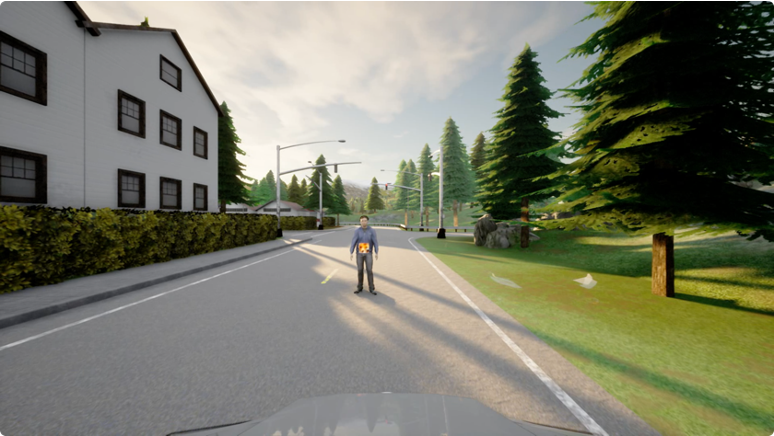} \\
			(a) truck scene & (b) mannequin scene \\
		\end{tabular}
	}
	\caption{Evaluation videos sticked with patches in final round}
	\label{fig: auto_videos}
\end{figure}

In the final round, we add 5 different scenes while the attack object is a mannequin simulated by Carla. And its scoring rules are the same as truck. The evaluation videos sticked with patches are shown in Figure \ref{fig: auto_videos}. We evaluated two patches of every team on each attacked object (truck and mannequin) respectively to compute the total score. The formula is summaried as:
\begin{equation}
	score=0.8*score_{truck}+0.2*score_{person}
\end{equation}
where $score_{truck}$ and $score_{person}$ denotes the score of truck and mannequin respectively. Finally, we ranked all the 10 teams based on the total score.

\subsection{Competition Results}

\begin{figure*}[t]
	\centering
	\resizebox*{\textwidth}{!}{
		\begin{tabular}{ccccc}
			\includegraphics[scale=0.5]{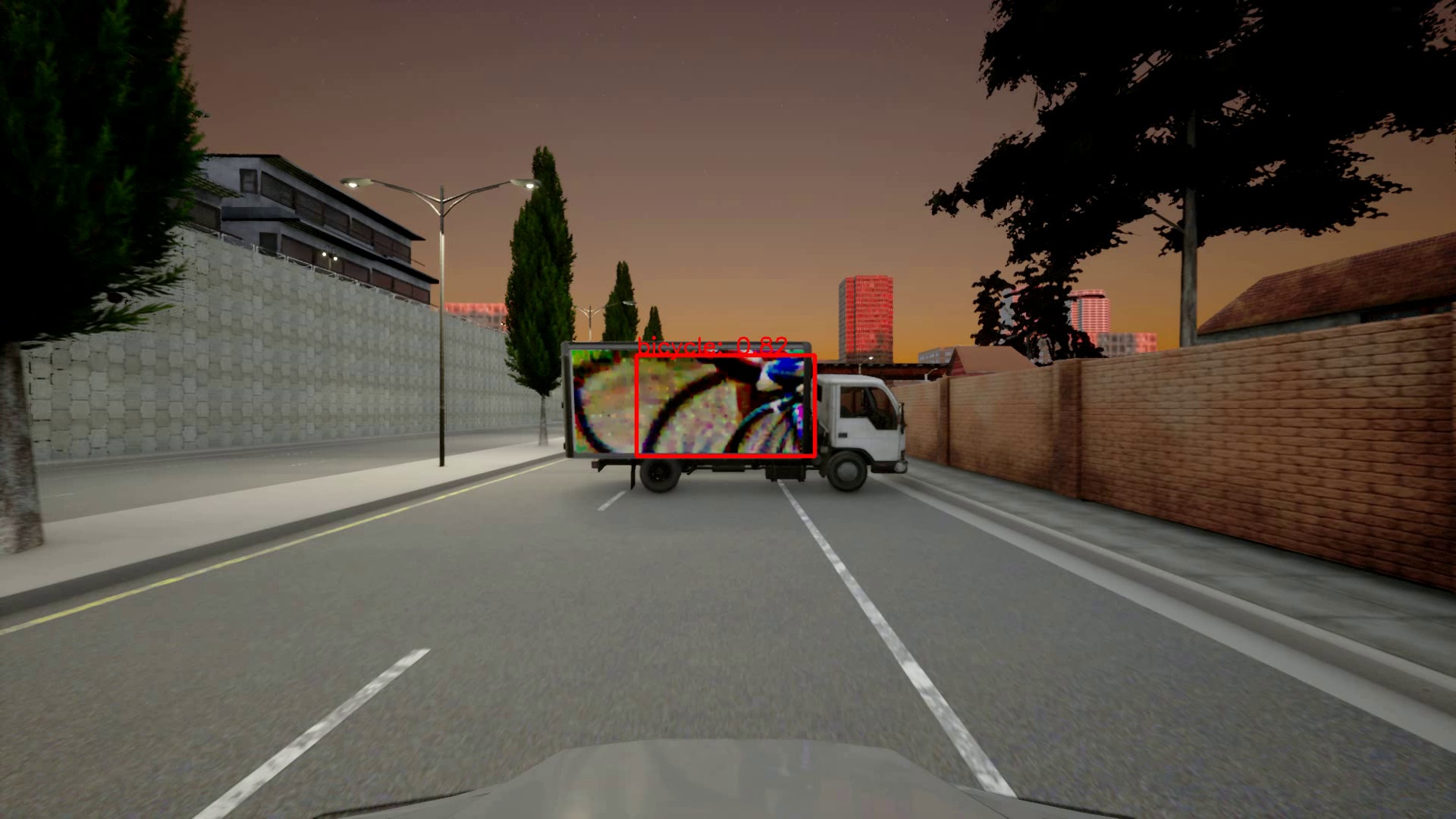} &
			\includegraphics[scale=0.5]{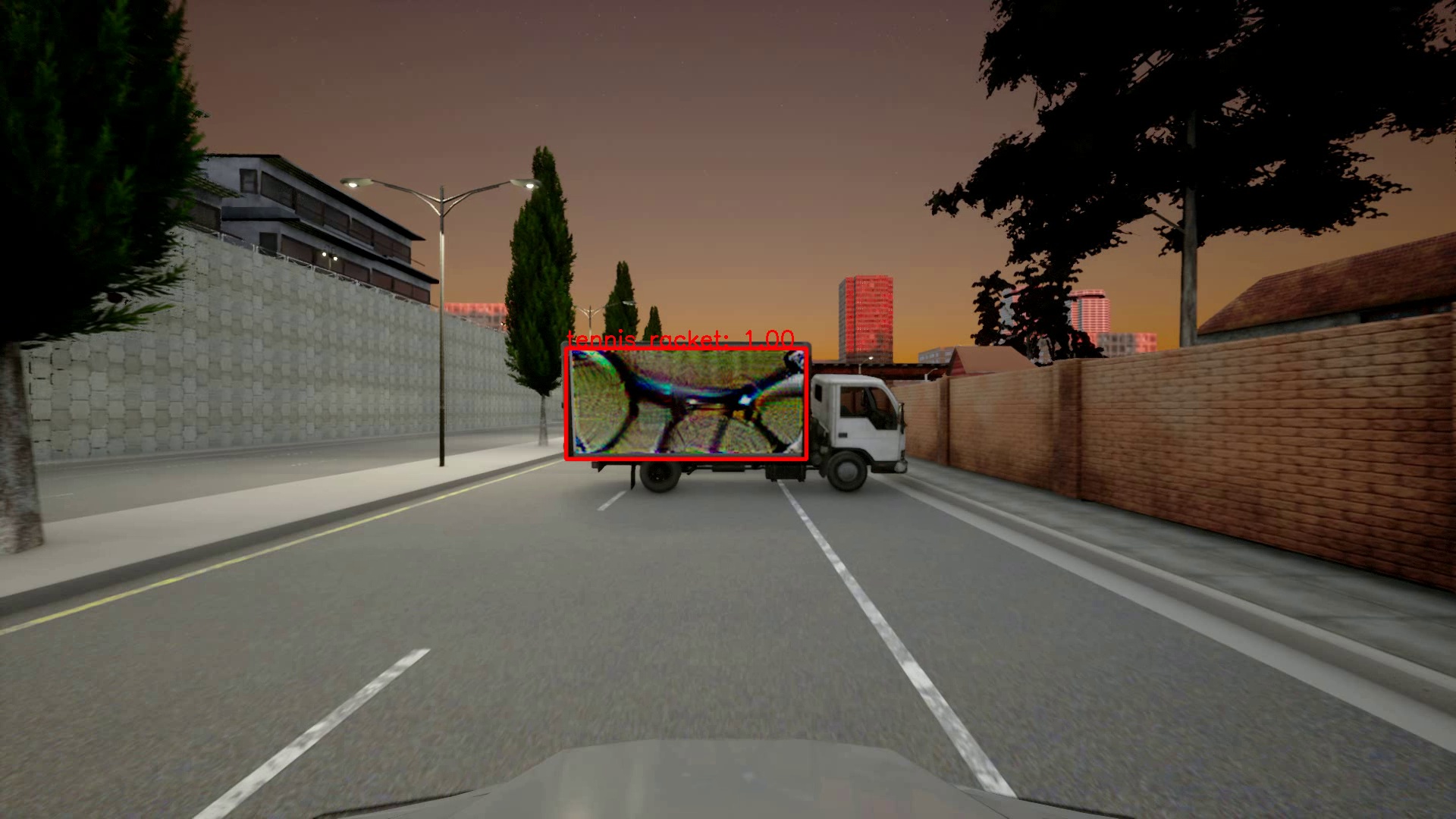} &
			\includegraphics[scale=0.5]{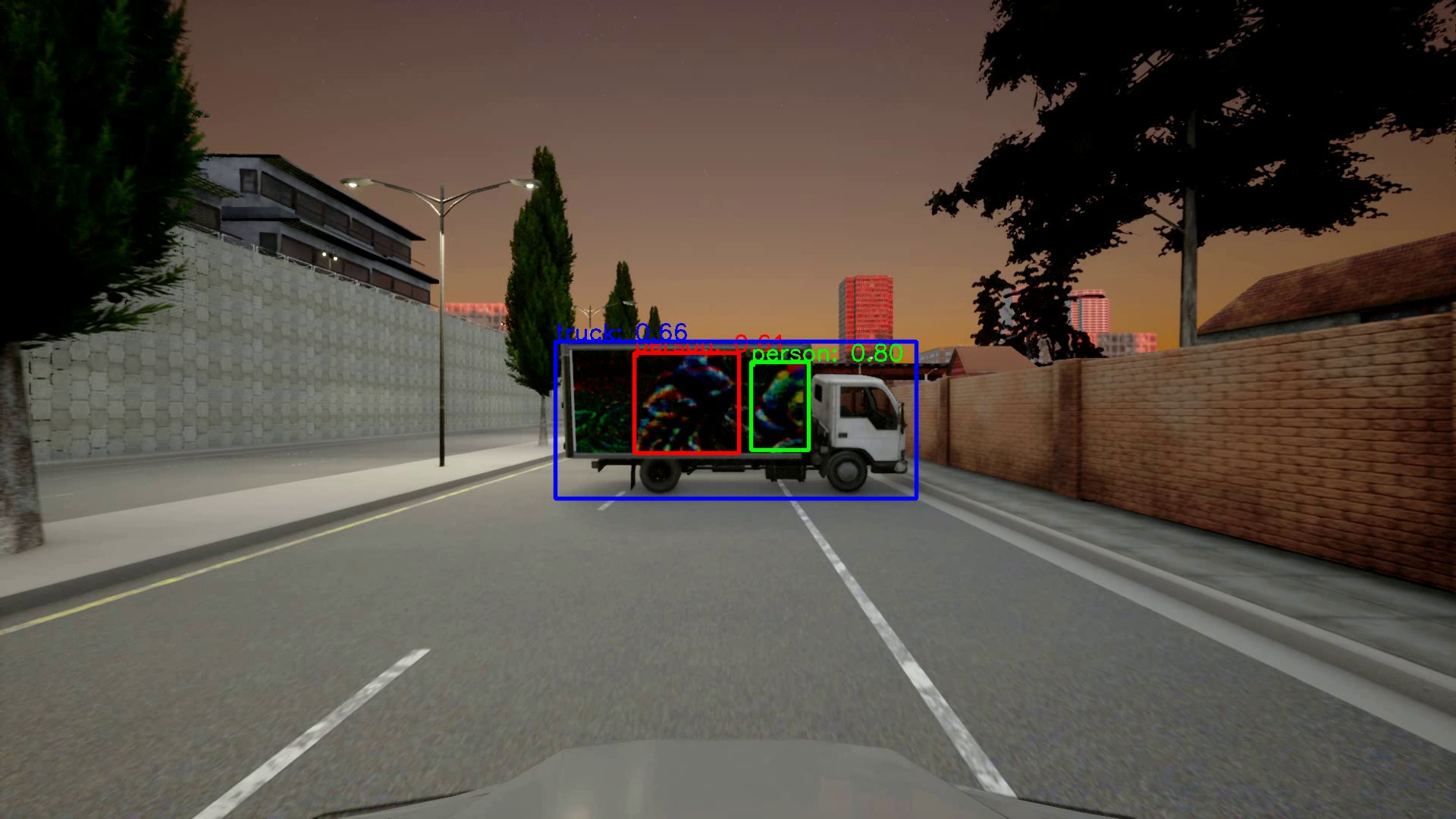} &
			\includegraphics[scale=0.5]{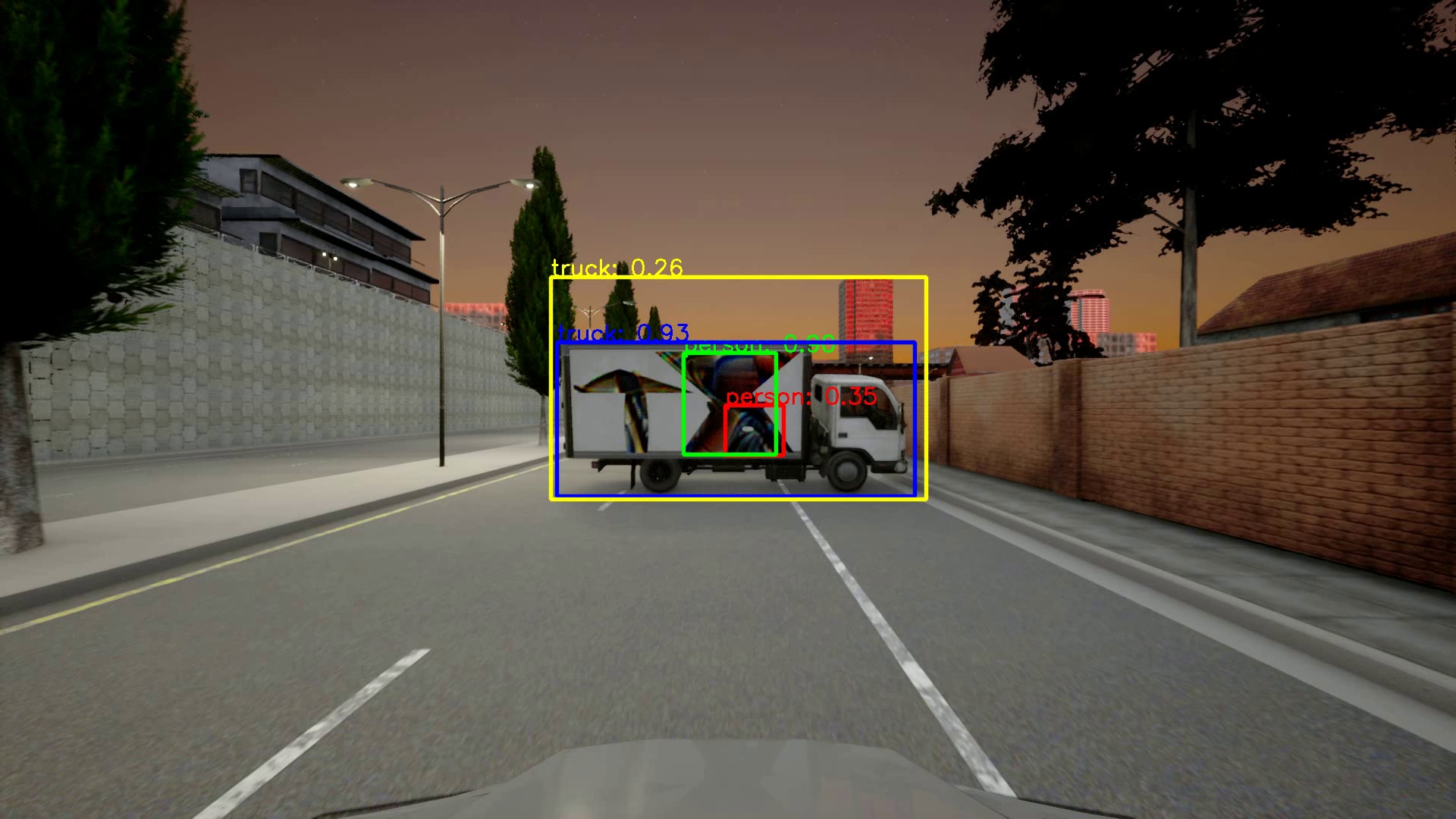} &
			\includegraphics[scale=0.5]{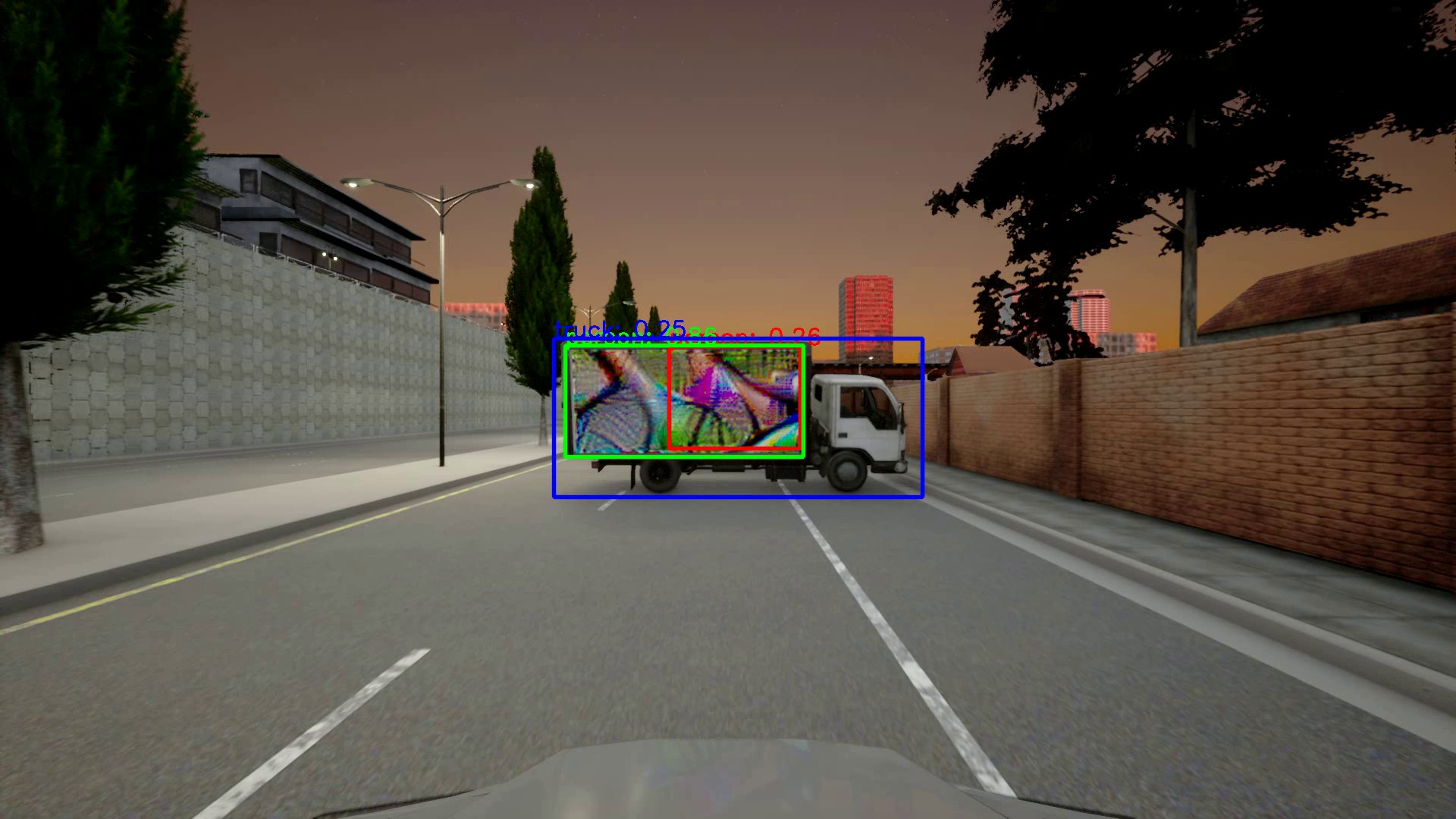} \\
			(a) 1st team & (b) 2nd team & (c) 3rd team & (d) 4th team & (e) 5th team \\
		\end{tabular}
	}
	\caption{The visual results of top-5 teams in final round on black-box detection model.}
	\label{fig:final-results}
\end{figure*}

In this challenge, there are totally 96 registered teams and throughout the preliminary round, we have received  about 15 valid results every day. Since the race started, we have recieved some meaningful solutions which have satisfying results not only on the white-box model but also on the black-box. The dectection model will fail to detect the right object class sticked with adversarial patch. The visual results are shown in Figure \ref{fig:final-results} and the quantitative results of top-5 teams are shown in Table \ref{tab:top-5-results}.
	
\begin{table}[h]
	\centering
	\caption{The quantitative results of top-5 teams in final round.}
	\resizebox*{0.4\textwidth}{!}{
		\begin{tabular}{c|ccc}
			\toprule
			Team Name & Truck & Mannequin & Total \\
			\midrule
			BJTU-ADaM & 0.79 & 0.00 & 0.63 \\
			XJTU\_Vanish & 0.67 & 0.19 & 0.57 \\
			YouOnlyAttackOnce & 0.45 & 0.20 & 0.40 \\
			XJTU-AISEC & 0.38 & 0.20 & 0.34 \\
			CETC-NHY & 0.39 & 0.00 & 0.31 \\
			\bottomrule
		\end{tabular}
	}
	\label{tab:top-5-results}
\end{table}

\subsection{Top Scoring Submissions}

\subsubsection{1st place: BJTU-ADaM}

\begin{center}
    \textbf{Team Member}
\end{center}

Zhiyu Lin, Yifei Gao, Shuang Li, Haonan Wang, Jitao Sang

\begin{center}
    \textbf{Method}
\end{center}

\begin{figure}[h]
    \centering
    \includegraphics[scale=0.8]{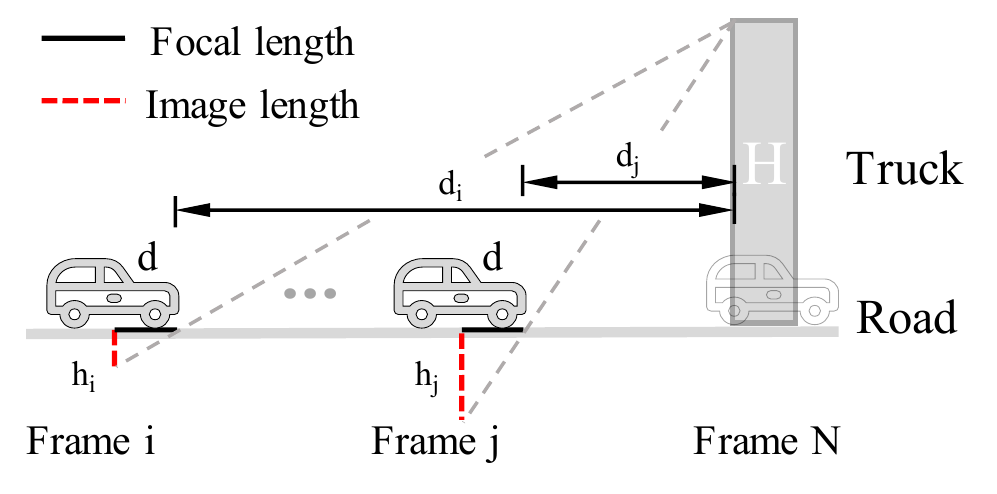}
    \caption{Pinhole imaging model for coordinate annotation}
    \label{fig: PIP-model}
\end{figure}

\textbf{Data annotation}: Inspired by the principal of pinhole imaging, we build a relationship between truck coordinate and the frame order, which substantially reduces the annotated cost. The framework is shown in Figure \ref{fig: PIP-model}.

Let $d$ denotes the focal length of viewfinder and $H$ denotes the height of truck (all constant values). Let $h_{i}$ denotes the image at frame $i$. Then we have $Hd=h_{i}d_{i}=h_{j}d_{j}$, where $d_{i}$ is the distance between the viewfinder position at frame $i$ and truck, which can be estimated by the frame order gap between frame $N$ (hyper-parameter) and frame $i$. Table \ref{tab:Hyper-parameters} shows the value of $N$ in each video.

\begin{table}[h]
    \centering
    \caption{Hyper-parameters of proposed model}
    \begin{tabular}{c|cccccccccc}
        \toprule 
        - & Video1 & Video2 & Video3 & Video4 & Video5 & \\
        \midrule
        
        N  & 269 &	283 & 285 &	282 &	298 \\
        \bottomrule
    \end{tabular}
    \label{tab:Hyper-parameters}
\end{table}

We assume that car moves with a constant speed, thus the zoom factor of two frames can be denoted as $\alpha_{ij}= h_{i}/h_{j}=d_{j}/d_{i}$. In this way, we can calculate the coordinates information of other frames through the coordinate information in the first frame and the zoom factor $\alpha_{0j}, j\in\{1,2,\cdots 231\}$.

\textbf{Adversarial patch training}: The core idea of our method is to update an adversarial patch $\delta$ through all video frames. Inspired by\cite{thys2019fooling}, we consider the following mathematical formulation to find an adversarial patch:
\begin{equation}
\mathop{\arg\min_{\delta}} \ \mathbb{E}_{x\in \mathcal{D}}[\mathcal{J}(A(x,\delta))]
\end{equation}
where $\mathcal{D}$ is data distribution over all video frames. We first interpolate adversarial patch $\delta$ to fit the frame-specific size of truck by utilizing the zoom factor, and stick it to the truck according to the coordinate information in data annotation stage.

We consider the anchor based detection method in this competition, where the output of each anchor point contains object probability $p_{obj}$ and predicted class $k$. For ensuring the stability of training process and the detection performance outside the attacking requirements, we filter out anchor points which satisfy: (1) $p_{obj}$ is larger than a given threshold $\alpha$, (2) class prediction $k$ belongs to the set of attack classes.

In order to solve the challenge of multi-class attack task, we attack the anchor points into specific target class $t$ by calculating cross entropy loss $\mathcal{L}_{cls}$. Furthermore, to ensure the immunity to noise and enhance semantic properties in real-world scenarios\cite{thys2019fooling}, the variation of the $\delta$ is considered as a constraint term $\mathcal{L}_{tv}$ in loss function. Overall, we calculate their summation scaled by a hyper-parameter $\beta$ as follows:
\begin{equation}
\mathcal{J} = \mathcal{L}_{cls} + \beta \cdot \mathcal{L}_{tv}
\end{equation}

\begin{center}
    \textbf{Submission Details and Results}
\end{center}

We leverage the detection model trained on COCO\cite{lin2014microsoft} dataset to generate patch. In the training process, the target class $t$ is set as a fixed class, an optimizer is used with an initial learning rate. To tackle with the black box attack, we choose YOLO\cite{redmon2016you} series algorithms and Faster R-CNN\cite{ren2015faster} for ensemble learning. The preliminary score is 0.74, the final score is 0.63. All scores are ranked first.

\subsubsection{2nd place: XJTU\_Vanish}

\begin{center}
    \textbf{Team Member}
\end{center}

Changfeng Sun, Haoliang Han, Xuchong Zhang

\begin{center}
    \textbf{Method}
\end{center}

\textbf{Overall framework.} The goal of the competition is to make the object detection models fail to correctly detect the vehicle.
The key idea of our method is to reduce the confidence of car, truck, and bus objects by attacking the output layer of detectors, thus make detectors unable to predict the correct bounding box.
We adopt~\cite{thys2019fooling} as the baseline method and make several improvements. 
Our total loss function can be expressed as follows:

\begin{equation}
    L = \alpha L_{adv} + \beta L_{TV} + \gamma L_{NPS}
\end{equation}
We take the sum of the three losses scaled by factors $\alpha$, $\beta$ and $\gamma$, and use the Adam~\cite{kingma2014adam} algorithm to optimise. The losses and detailed improvements to the method are described in the following content.

\textbf{Attack environment simulation. }
In order to conduct the proposed attack method, it is necessary to construct a visual environment to simulate the real world. 
We adopt the driving videos collected in the Carla simulation platform provided by the organizer to construct the visual environment.
These videos include 5 different road scenes used in the preliminary competition and we extract frames from these videos with 24 FPS.
Finally, 1150 simulated images are obtained and we evaluate the attack method based on theses images.
To meet the need of adding adversarial patch upon the wagons, the location of the wagons in different images is necessary.
We use a software\footnote{https://github.com/wkentaro/labelme} to label location using two pixel points $ (x_1, y_1),(x_2, y_2)$ which represent the upper left corner and the lower right corner of the bounding box of the wagons.
Based on this, we establish a mask matrix to indicate the pixels which would be replaced by the patches.

\textbf{Resize.} 
Because the wagons are different across the different images, adjusting the patch to fit the different wagons is a key problem to conduct attack. 
Through the observation of the simulation videos, we found that, in different images, the wagons do not appear great rotation, occlusion, deformation and other problems, only the size changes much due to the distinct distance.
Therefore, we just create an initial patch and resize it to simulate the change of the patch. 
It is worth mentioning that the size of the original patch plays an important role to the attack success rate.
If the size is too big, the patch would lose a lot of pixels when it is resized to a small one, thus the gradient can not be fed back to the lost pixels and resulting in the failure.
If the size is too small, the patch would have little room for change, thus the attack would be more difficult.

\textbf{Joint training.}
In order to enhance the transferability of the adversarial patch between different models, we introduce a joint training strategy. Intuitively, adversarial patches are more transferable between detection models with the same backbone because they extract features in a similar way. The transferability of object detection models with different backbones would be weaker. Some of our experiments have also demonstrated this. We selected three object detection models
for joint training. The joint loss function  $L_{adv}$ can be written as follows:
\begin{equation}
    L_{adv}=\lambda_{1}L_{1}+\lambda_{2}L_{2}+\lambda_{3}L_{3}
\end{equation}
where $\lambda_{1}, \lambda_{2}$ and $\lambda_{3}$ are the hyper-parameters.

\textbf{Physical realizability.} To ensure physical realizability of the adversarial patch, we follow~\cite{sharif2016accessorize} to minimize total variation (TV) and the non-printability score (NPS) of it. For a adversarial patch $p$, the calculation of total variation loss can be written as follows:
\begin{equation}
    L_{TV}=\sum_{i,j}\sqrt{((p_{i,j}-p_{i+1,j})^2+(p_{i,j}-p_{i,j+1})^2)}  
\end{equation}
where $p_{i,j}$ is the pixel value of $p$ at coordinate $(i, j)$.

To make the adversarial patch generated in the simulation environment can be printed correctly, we need to ensure that the colour is in the set of printable colours. We utilize the non-printability score (NPS) to constrain it. This loss can be calculated as follows:
\begin{equation}
    L_{NPS}=\sum_{p_{patch}\in p} \min_{\Tilde{p}_{print}\in p} \left | p_{patch}-\Tilde{p}_{print} \right | 
\end{equation} 
where $p_{patch}$ is a pixel in of the adversarial patch $p$ and $\Tilde{p}_{print}$ is a colour in a set of printable colours $\Tilde{p}$.

\begin{center}
    \textbf{Submission Details and Results}
\end{center}

\textbf{Submission Details:}
We iterate 20 times to update the original patch. 
The preliminary score is 0.691667, the final score is 0.571140788. All scores are ranked second.

\subsubsection{3rd place: YouOnlyAttackOnce}

\begin{center}
    \textbf{Team Member}
\end{center}

Zeyu Liu, Yuanzhe Pang, Chengqi Duan

\begin{center}
    \textbf{Method}
\end{center}

Our work consists of two parts, namely the preprocess of data and the generation of an adversarial patch.

    To process data, we first split the videos into images and generate differential images between the patched and unpatched images. We then utilize cv2 to identify every rectangle in the differentiated image which lead us to the body of the truck where we place our patches.

    In order to mislead a detector so that it cannot identify cars, buses and trucks in the real world, we formulate our attack through a two-step approach. The first step is to mislead the detector in the digital world, and the second to transfer it to the physical world.

    In the first step, we consider an untargeted approach and hide our target object through discrediting the object's objectiveness. 
    For a surrogate white box model YOLOv3, we define its objectness score as $O$, class prediction $C$ and set of targets to suppress $S$. Therefore, we can formulate the untargeted attack as minimizing the above loss function:
    
    \begin{center}
        $l_{\text{obj}} = \sum \limits_{i, \mathop{\arg \max} C_i \in S} O_i$
    \end{center}

    In addition, we use momentum \cite{https://doi.org/10.48550/arxiv.1710.06081} to update our patch in order to boost transferability. 

    In the second step, we consider transfering our attack to the physical world. This includes decreasing feature loss when printing our patch and adapting it to real world light and environmental changes. This can be solved by introducing regularization terms including TV-loss and the non-printablility score\cite{thys2019fooling}. 
    
    TV loss is introduced to decrease feature loss when transfering our patch to the physical world. This ensures that our patch has smooth colour transitions rather than noisy and grained ones.
    \begin{center}
    $l_{\text{tv}} = \sum \limits_{i, j} ((x_{i, j +1} - x_{i, j})^2 + (x_{i +1, j} - x_{i, j}))^2$
    \end{center}
    where $x_{i,j}$ is the pixel in $i^{th}$ row and $j^{th}$ column. 
    
    Non-printability score is a factor measuring how well colors can be printed in reality. Given a set of printable colors $C$, NPS is defined as:
    \begin{center}
    $l_{\text{nps}} = \sum \limits_{i, j} \min \limits_{c_{\text{print} \in C}} |x_{i, j} - c_{\text{print}}|$
    \end{center}
    And with that the total loss can be formulated as:
    \begin{center}
    $L = l_{\text{obj}} + \mu_1 l_{\text{tv}} + \mu_2 l_{\text{nps}}$
    \end{center}


\begin{center}
    \textbf{Submission Details and Results}
\end{center}

 With these measures, we achieve 0.59 points out of 1 on a white box and black box validation model. And our scores are ranked third place in final stage.

\subsubsection{4th place: XJTU-AISEC}

\begin{center}
    \textbf{Team Member}
\end{center}

Chen Ma, Junhao Zheng, Yijia Li, Chao Shen, Chenhao Lin

\begin{center}
    \textbf{Method}
\end{center}

Our idea has two main parts: mask generation and adversarial patch generation.

\textbf{Simulation of the mapping process:} To reduce the huge time cost of the simulator rendering and shooting process as a result caused, we extract scene video frames and paste patches in the target area as an alternative method. Since vehicle speed is stable, we sample images every nine frames and obtain the coordinates of the target area using polynomial fitting. Image transformation based on the coordinates allows the patch to be placed precisely on the target area in each frame.

\textbf{Mask generation:} This process is divided into three steps: initial mask generation, reprocessing and screening. \textbf{(1) Preliminary formation of the mask.} Considering that the shape of the mask and the number of connected domains are not fixed, our method is: firstly, we design a variety of shape and size adjustable graphic blocks (such as triangles, ellipses, rectangles, etc.), and then randomly place several graphic blocks on a blank mask as effective area of the mask. In order to evaluate the effectiveness of the mask, we take a pure black background as adversarial patch pattern, paste the adversarial patch in target area on the extracted images, and count attack success rate. \textbf{(2) Reprocessing of mask.} In order to obtain more effective masks, we select the mask with higher attack rate among the initially generated masks and continue to superimpose graphic blocks to derive more masks. This time the superimposed graphic blocks may be black or white, representing addition or elimination of the effective area. \textbf{(3) Screening of masks.} In order to limit the size of adversarial patch area, we perform the final screening of masks and removed the masks whose effective area exceeds a certain threshold. Finally, We select the best mask in remaining masks.

\textbf{Adversarial patch generation:} We use the mask selected by the above method to generate adversarial patch.
Then we use the improved algorithm to carry out optimization attack. The basic process is as follows: An appropriate initial pattern of adversarial patch is set, and the patch is attached to the target area according to the mask shape in each frame of picture, and the patch pattern is iteratively optimized according to the designed loss function. \textbf{(1) Adversarial patch processing.} As mentioned earlier, patches need to be perspective transformed. In order to simulate the texture effect more realistically, we apply the Gaussian filtering to the patch before and after the perspective transformation, which can make the patch closer to the rendering effect in physical scene. 
\textbf{(2) Design of loss function.} For example, if the model fails to detect "car", "truck" and "bus", the attack is considered successful. To simplify the description, we refer to "car," "truck," and "bus" as "attack classes," and the remaining as "other classes." We use the detection results before NMS processing to calculate the loss function, and the formula is as follows:
\[x_{adv} = M \odot T_2(\sigma) + (1-M) \odot x\]
\[arg \mathop{\min}_{\sigma} \mathbb{E}_{x \sim X} [J_{obj}(T_1(x_{adv}), y_{attack})] \]
where X is the dataset, $T_2$ is a distribution over patch transformations and $\odot$ denotes element-wise product. The loss function $J_{obj}$ represents the difference of the maximum probability of “attack classes” and the maximum probability of "other classes". 

\textbf{Ways to improve robustness.}\textbf{ (1) Shuffling order of the dataset.} Since the dataset are continuous video frames, we thus randomly shuffle the dataset for each training round to eliminate the adversarial patch bias caused by the distribution between different batches. \textbf{(2) Adding perturbation.} We add changed small amplitude perturbations to improve robustness.
\textbf{(3) Two image pre-processing methods.} Considering the two dominant image pre-processing approaches used by object detection models, 
pre-processing methods are implemented and target detection is carried out respectively in the process of generating adversarial patches, so that the generated adversarial patches have attack effect on both pre-processing methods.

\begin{center}
    \textbf{Submission Details and Results}
\end{center}

In the optimization process, the batch size is set as 15,
and the number of iteration rounds of adversarial patch finally submitted is 90. Our team scores 0.567413 point in the preliminary judging and 0.342597 point in the final, ranking fourth.

\subsubsection{5th place: CETC-NHY}

\begin{center}
    \textbf{Team Member}
\end{center}

Enhui Xu, Jincai Xu, Shu Xu

\begin{center}
    \textbf{Method}
\end{center}

Our solution is mainly derived from this work~\cite{thys2019fooling}, where the optimization goal consists of the following three parts:

$L_{n p s}$ The non-printability score~\cite{sharif2016accessorize}, a factor that represents how well the colours in our patch can be represented by a common printer. Given by:
\begin{equation}
L_{n p s}=\sum_{p_{\text {patch }} \in p} \min _{c_{\text {print }} \in C}\left|p_{\text {patch }}-c_{\text {print }}\right|
\end{equation}
Where $p_{\text {patch }}$  is a pixel in of our patch $P$ and $c_{\text {print }}$ is a
colour in a set of printable colours $C$. This loss favours colors in our image that lie closely to colours in our set of printable colours.

$L_{t v}$ The total variation in the image as described in~\cite{sharif2016accessorize}. This loss makes sure that our optimiser favours an image with smooth colour transitions and prevents noisy images. We can calculate $L_{t v}$ from a patch $P$ as follows:
\begin{equation}
L_{t v}=\sum_{i, j} \sqrt{\left(\left(p_{i, j}-p_{i+1, j}\right)^2+\left(p_{i, j}-p_{i, j+1}\right)^2\right.}
\end{equation}
The score is low if neighbouring pixels are similar, and high if neighbouring pixel are different.

$L_{o b j}$ The maximum objectness score in the image. The goal of our patch is to hide truck or persons in the image. To do this, the goal of our training is to minimize the object or class score outputted by the detector. 

Out of these three parts follows our total loss function:
\begin{equation}
L=\alpha L_{n p s}+\beta L_{t v}+L_{o b j}
\end{equation}

We take the sum of the three losses scaled by factors $\alpha$ and $\beta$ which are determined empirically, and optimise using the Adam algorithm.

The goal of our optimizer is to minimise the total loss $L$. During the optimisation process we freeze all weights in the network, and change only the values in the patch. The patch is initialised on random values at the beginning of the process.

\begin{center}
    \textbf{Submission Details and Results}
\end{center}

The target detection model used for training is YOLOv3. The video data provided by the competition is extracted and labeled, and the images are resized to (640,384) and fed into the model for training.
And at last we got fifth place at last.

\section{Face Recognition Security Competition}
\label{sec:4}
This competition track focuses on the security of face recognition models. As one of the most mature technologies in the field of computer vision, face recognition is widely used in multiple high-value sensitive scenarios such as mobile payment, security systems, and self-service. However, the current face recognition models based on deep learning algorithms are vulnerable to adversarial example attacks. That is, the system can be misled to make wrong predictions by making small changes to the input face data, which seriously threatens the security of face recognition applications.

This competition track simulates adversarial attacks under the face verification scenario, in order to promote the research on safe and controllable face recognition algorithms. To approximate physical-world attacks, this track adopts adversarial patches as the attacking patterns. Under the black box scenario where the model details are unknown, the participants need to generate adversarial patches to make the face verification system recognize the adversarial face image as the specified target identity. The submissions are ranked based on the attack success rate. This competition track aims to discover more stable attack algorithms for evaluating the security of face recognition models and consequently facilitate the development of more robust face recognition models.

\subsection{Competition Rules}

This track is divided into three stages: preliminary competition, preliminary evaluation and final competition. The rules are as follows.

\textbf{Preliminary competition:} We adopt the famous face recognition dataset LFW, containing 3000 pairs of aligned face images (x{1}, x{2}) belonging to two different identities. For each pair, the participants need to generate an adversarial patch on (x{1}) and obtain the adversarial example x{adv}. We input x{adv} and x{2} to the face verification model. If the model wrongly recognizes x{adv} and x{2} as the same identity, the attack is regarded successful. We will score the submissions based on the attack success rate. In this stage, we adopt three black-box face recognition models, i.e., the participants cannot access the models used for evaluation. The total number of adversarial patches for each image should be no more than 5, and the area of these patches should not exceed 10\% of the original image area.

\textbf{Preliminary evaluation:} The last submissions at the end of the preliminary competition will be evaluated on three different black-box face recognition models. At the same time, the participants need to email the code and algorithm description corresponding to the last submission. The experts of the organizing committee will evaluate the participants' plans and judge whether the adversarial patches generated by the participants can be realized in the physical world. If they pass the expert review, they will be ranked based on the attack success rate, and the top 10 teams will enter the final competition.

\textbf{Final competition:} The participants need to attend the Zhongguancun Forum offline and carry out physical-world attacks against commercial face recognition systems (including live detection). Given the target identities, the participants need to take their generated adversarial patches to the forum, deceive the provided commercial face verification model within the required time, and are finally ranked based on the attack success rate. At this stage, the adversarial patches need to satisfy two requirements: 1. The area of its frontal part does not exceed 60 square centimeters. 2. It can pass the live detection model of the system.

\textbf{Data:} We adopt 3000 pairs of face images from the LFW dataset. Each image is processed by MTCNN for face detection and alignment, and then resized to the resolution of 112*112 of .png format. The participants need to add adversarial patches to images named xxxx.png and make them be recognized as the same identity as xxxx\_compare.png. In the preliminary round, the participants need to generate 3000 adversarial samples named as 0001.png, 0002.png, ..., 3000.png, which are compressed into .zip format and then submitted. The adversarial sample needs to be an 8-bit RGB image with the same size as the original image (i.e. 112*112 pixels).

\textbf{Evaluation Metric:} In the preliminary competition, we score the submissions based on the attack success rate (higher is better), as
\begin{equation}
    Score=\dfrac{1}{3}\sum_{i=1}^{3}\dfrac{1}{3000}\sum_{j=1}^{3000}I\left(M_{i}\left(x_{adv}^{j}, x_{2}^{j}\right)\right)
\end{equation}
where $I$ is the indicator function, $M_{i}$ is the $i\-th$ face verification model which outputs 0 or 1 to represent whether $x_{adv}^{j}$ and $x_{2}^{j}$ belong to the same identity, used to determine whether the attack is successful. Furthermore, if the adversarial example does not satisfy the following requirement, the attack is considered unsuccessful.

\textbf{Connected Domain Requirement:} In evaluation, we subtract the adversarial example submitted by the participants from the original image to get the added perturbation, and then calculate the number and size of the connected domains in the perturbation. A connected domain is regarded as an added adversarial patch, and the number of adversarial patches is required to be no more than 5, and the sum of the area of adversarial patches does not exceed 10\% of the original image pixels (i.e., 112*112*10\%=1254).

\subsection{Competition Results}

In this competition track, there are 178 teams participate in the preliminary competition, and submit results 1821 times totally. We evaluate the last submissions of all teams then select the top 10 teams ranked by attack success rate as winners enter the final competition. After a fierce competition, 2 teams break out from all the 10 teams and obtain all the scores of all tasks, so they both are the fist winner.
\begin{table}[h]
	\centering
	\caption{The score of top-2 teams in final competition.}
	\resizebox*{0.26\textwidth}{!}{
		\begin{tabular}{c|ccc}
			\toprule
			Team Name & score \\
			\midrule
			TianQuan $\&$ LianYi & 100 \\
			Deep Dream & 100 \\
			\bottomrule
		\end{tabular}
	}
\end{table}

\subsection{Top Scoring Submissions}

\subsubsection{1st place: TianQuan $\&$ LianYi}

\begin{center}
    \textbf{Team Member}
\end{center}

Yu Wu, Jian Lin, Tianpeng Wu, Ye Wang and Yu Fu

\begin{center}
    \textbf{Method}
\end{center}

We adopt Adv-Markup~\cite{yin2021adv} as the backbone architecture, which can achieve good physical attack results on certain white-box FR (face recognition) models. To achieve better attack robustness over black-box commercial FR platforms, we developed the following several improvements upon it.

\textbf{Design of face mask:} The face mask domain can be set to auto landmarks or hand-crafted design. Theoretically speaking, the attack of FR models will work better with the larger area of face mask. However, under real-world application scenes, a large area of adversarial mask is unable to pass the face anti-spoofing.
In our study, a simple connected region including eyebrows, eye sockets, nose and lips in a limited area is chosen to generate adversarial mask. When the attacker wear the generated physical face mask, it can be also work for silent face anti-spoofing.

\textbf{Number of white-box models:}
To improve generalization ability of the method, we added several popular white-box face recognition models for training and validation, including MobileFaceNet~\cite{chen2018mobilefacenets}, GhostNet~\cite{han2020ghostnet}, the series of Arcface~\cite{deng2019arcface}, the series of PartialFC~\cite{an2021partial}, the series of Cosface~\cite{wang2018cosface}, the series of Magface~\cite{meng2021magface},  and the series of Adaface~\cite{kim2022adaface}. 

\textbf{Strategy of meta-training and meta-testing:}
The increased number of face recognition models improve the robustness but causes  the larger memory occupation, we designed a dynamic learning strategy to use more models with lower memory occupation. 
For instance, there are total 20 white-box models, where 4 models are used for meta-training, and the remaining models are used as the black-box testing models. The meta-testing models are further divided into 4 groups. In each epoch, one training model randomly matches a set of meta-testing models. Therefore, meta-training models and meta-testing models will be dynamically adjusted during training.

\textbf{The weight of attack loss increases:}
In Adv-Markup, the weights of different losses are 1. Since we adjusted the training strategy and the number of models in meta-learning, we finally used weighted function.
In our final experiments, the success rate of an attack over white-box face recognition models was associated with an increase in the weight of attack loss and the number of iterations.
\begin{center}
    \textbf{Submission Details and Results}
\end{center}

\textbf{Submission details:} The input size of the network is $500 \times 500$, and the final mask domain is $360 \times 320$. The number of iterations is 75,000 in the final. There are 300 images for both the target and source person. In each iteration, we randomly chose one image for each person as the input. The weight of attack loss is 14 and the others are 1. The final number of the white-box FR models is 18.

\textbf{Results:} The final score is 100, which is ranked first.

\subsubsection{1st place: Deep Dream }

\begin{center}
    \textbf{Team Member}
\end{center}

Huipeng Zhou, Yajie Wang, Yuhang Zhao, Shangbo Wu, Haoran Lyu

\begin{center}
    \textbf{Method}
\end{center}

To obtain a higher success rate of black-box targeting attacks on unknown face recognition models. We followed the suggestion of ~\cite{Xiao2021ImprovingTO} and first used the GenAP-DI~\cite{Xiao2021ImprovingTO} method to obtain a more suitable initialisation, and then used other gradient transferability-based methods (e.g. MI-FGSM~\cite{Dong2018BoostingAA}, TI-FGSM~\cite{Dong2019EvadingDT}, DI-FGSM~\cite{Xie2019ImprovingTO}, etc.) to optimize the adversarial perturbation further. In this competition, we ensemble the robustness model h non-robustness model and set different weight shares for different models. To avoid the problem of overfitting of local surrogate models leading to lower transferability, we break the static order of execution between multiple methods to dynamic ones. We also explore the effect of factors such as step size, perturbation size, and different Masks on transferability. The next section explains our method in more detail.

\begin{center}
    \textbf{Submission Details and Results}
\end{center}

\begin{figure}[]
    \centering
    \includegraphics[width=0.45\textwidth]{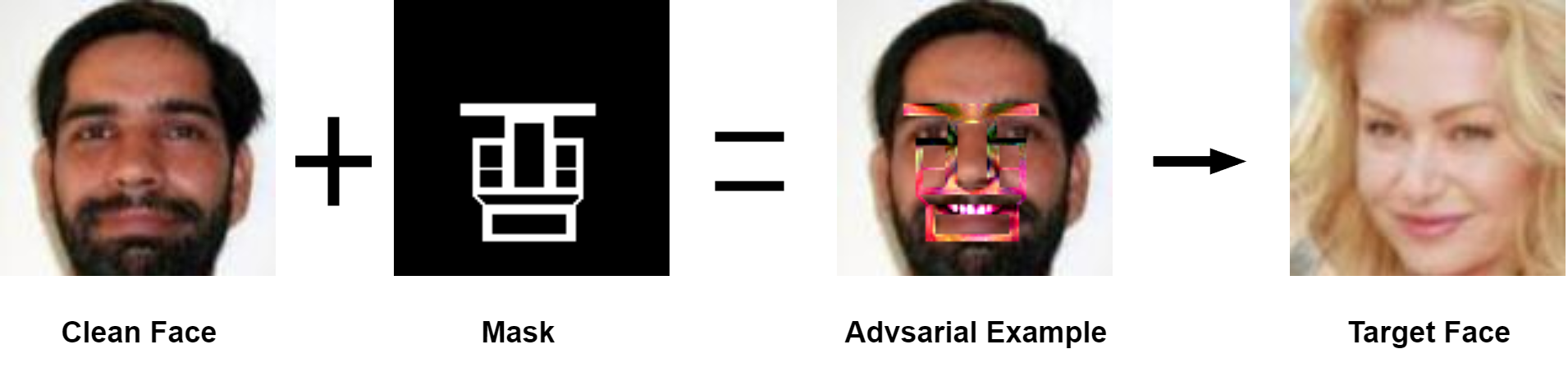}
    \caption{Visualisation of adversarial examples generated using our method on the LFW~\cite{Huang2008LabeledFI} dataset. The adversarial perturbations we generate are evenly distributed over the face, rather than over the five senses (e.g. eyes, nose, etc.).}\label{vis_adv}
\end{figure}

\textbf{Mask.} As the number of adversarial patches is required to be no more than five, the total area of the patches does not exceed 10\% of the pixels of the original image. By considering the in live verification aspect of the face recognition task, we target the region of the adversarial perturbation on the facial region rather than on the five senses. We carefully designed the area of the mask, and the adversarial example we generated is shown in Figure \ref{vis_adv}.

\textbf{Surrogate models.} We ensemble 12 face recognition models as local surrogate models (4 robust and 8 non-robust models). We chose cosine similarity as the loss function, and each model was given a dynamic weight share in the iterative attack, with a $1 - cosine  similarity$ rule.

\textbf{Method.} Our approach is divided into two parts. Part 1: We first initialize the adversarial perturbations using the GenAP-DI method using the latent space of StyleGAN2~\cite{Karras2020AnalyzingAI}. Given an original face $X$ , mask $M$, StyleGAN2 model $S$ , $W^{+}$ is the latent space representation vector of the original face $X$ in StyleGAN2. The acquired adversarial example $X_{adv}$ can be expressed as:
\begin{equation}\label{con:2}   
    X_{adv}=X \odot (1-M)+S(W^{+}) \odot M
\end{equation}
Part 2: To further improve the transferability of the adversarial examples, we take the adversarial example $X_{adv}$ generated in part 1 and optimise it again. We ensemble the MI-FGSM, TI-FGSM, DI-FGSM, SI-FGSM~\cite{Lin2020NesterovAG}, and $S^{2}$I-FGSM~\cite{Long2022FrequencyDM} methods. For the order of execution of these methods, we performed dynamic random adjustments in the iterative attack.

\textbf{Submission Results.} The final score is 0.634889 and 0.592222.

{\small
\bibliographystyle{ieee_fullname}
\bibliography{egbib}
}

\end{document}